# Why It Takes So Long to Connect to a WiFi Access Point

Changhua Pei[†], Zhi Wang[§], Youjian Zhao[†], Zihan Wang[§], Yuan Meng[†], Dan Pei[†*]
Yuanquan Peng[‡], Wenliang Tang[‡], Xiaodong Qu[‡]
[†]Tsinghua University    [‡]Tencent    [§]Graduate School at Shenzhen, Tsinghua University
[†]Tsinghua National Laboratory for Information Science and Technology (TNList)

*Abstract*— Today's WiFi networks deliver a large fraction of traffic. However, the performance and quality of WiFi networks are still far from satisfactory. Among many popular quality metrics (throughput, latency), the probability of successfully connecting to WiFi APs and the time cost of the WiFi connection set-up process are the two of the most critical metrics that affect WiFi users' experience. To understand the WiFi connection set-up process in real-world settings, we carry out measurement studies on 5 million mobile users from 4 representative cities associating with 7 million APs in 0.4 billion WiFi sessions, collected from a mobile "WiFi Manager" App that tops the Android/iOS App market. To the best of our knowledge, we are the first to do such large scale study on: how large the WiFi connection set-up time cost is, what factors affect the WiFi connection set-up process, and what can be done to reduce the WiFi connection set-up time cost. Based on our data-driven measurement and analysis, we reveal the following insights: (1) Connection set-up failure and large connection set-up time cost are common in today's WiFi use. As large as 45% of the users suffer connection set-up failures, and 15% (5%) of them have large connection set-up time costs over 5 seconds (10 seconds). (2) Contrary to the state-of-the-art work, *scan*, one of the sub-phase of four phases in the connection set-up process, contributes the most (47%) to the overall connection set-up time cost. (3) Mobile device model and AP model can greatly help us to predict the connection set-up time cost if we can make good use of the hidden information. Based on the measurement analysis, we develop a machine learning based AP selection strategy that can significantly improve WiFi connection set-up performance, against the conventional strategy purely based on signal strength, by reducing the connection set-up failures from 33% to 3.6% and reducing 80% time costs of the connection set-up processes by more than 10 times.

## I. INTRODUCTION

In recent years wireless data traffic has seen an exponential rise due to the explosion of smart devices. Among these wireless networks, 802.11 wireless LAN (WiFi) has served a dominate fraction of today's wireless traffics. In the past decade, over 1 billion WiFi APs (Access Points) have been sold and deployed to provide wireless connectivity [1]. WiFi hotspots that are found everywhere today have been used even when users are using smart devices with 3G/4G cellular networks supported.

However, network performance and user experience in WiFi networks are still not satisfactory: based on our measurement studies over 5 million users using WiFi networks in urban areas, as large as 45% of mobile devices fail in establishing a WiFi connection with the corresponding APs and 15% (5%) of successful WiFi connections suffer connection set-up time costs over 5 seconds (10 seconds). It is thus critical to understand the WiFi connection set-up process, the first step to use a WiFi network, including how the WiFi connection set-up performance is, why there is a significant fraction of connection failures and large connection set-up time cost events, and what can be done to reduce the WiFi connection set-up time cost.

Previous measurement studies on WiFi networks have been focused on general user experience metrics (*e.g.*, bandwidth and latency experienced in WiFi networks) and few focus on the performance of WiFi connection set-up process. [2] gives the first try to call people's attention on the connection set-up time cost. They collected data from a handful of voluntary Android smartphones in controlled environment and observed that the large connection set-up time cost is mainly caused by the loss of DHCP packets. However, the performance of WiFi connection set-up processes in the wild still remain unknown and there lacks thoroughly investigation in a larger scale. In our paper, we aim to address this problem. We collected connection log data from millions of mobile devices which are equipped with a popular Android App called "WiFi Manager". Firstly, the "WiFi Manager" App enables us to break down the connection set-up process to sub-phases to find some observations which are not discovered before. Secondly, the large-scale data collection enables us to train a machine learning model which can help us to predict the connection set-up time cost right before the user's connection attempt. This model can be further used to help mobile users select the best APs to reduce the connection set-up time costs.

Our contributions can be summarized as follows.

In the macro-level, we explore how the connection set-up time cost looks like in the wild. We continuously collect connection log from 5 million distinct mobile devices equipped with "WiFi Manager" App and collect 0.4 billion WiFi connection attempts. We find 45% of connection attempts fail. Among those successful connection attempts, 15% (5%) are larger than 5 seconds (10 seconds). Based on the **x-y** visualization and correlation analysis (*e.g.*, Relative Information Gain), we find that besides the well-known signal strength which affects the connection set-up processes, knowing the AP model and the mobile device model has great help to predict the connection set-up time cost. This is mainly because APs (mobile devices) always have similar characteristics if they are the same model. More detailed analyses can be found in §V.

---

* Dan Pei is the corresponding author.

In the micro-level, we break down the whole WiFi connection set-up process into sub-phases and calculate the fraction of time spent on each phase. We find that for more than half of the connection attempts whose total connection set-up time costs are larger than 15 seconds, 47% of the time is spent on *scan* phase. This is out of our expectation because the previous findings show that most of the time is spent on *DHCP* phase instead of *scan* phase [2]. By tracking the WiFi network state transitions of these connection set-up processes, we find anomaly transitions to *Disconnected* state cause the re-connecting. The further evidence shows that these are mainly caused by undesirable packet loss between the mobile devices and APs for the following reasons: variance of signal strength, WiFi interference, high load causing system response delay of mobile devices and APs.

At last, we show that all the above observations can be utilized to improve the performance of WiFi connection set-up process by helping mobile users to select better APs. It is not a good idea to purely use the signal strength measured at a mobile device to decide which AP to connect. We propose a machine learning based AP selection method by adding few easily obtained metrics as extra features including mobile device model, AP model, number of devices associated to the AP, hour in the day that a connection attempt happens. Using the dataset collected in the wild, we conduct what-if analysis to evaluate the performance gain after using our machine learning model to help mobile users select the best APs. The evaluation results show that the machine learning based AP selection method performs well compared to the baseline algorithm (selecting APs with the strongest signal strength). Our solution can reduce the ratio of connection failures from 33% to 3.6% and the 80% connection set-up time costs are reduced by 10×.

The rest of the paper is organized as follows: In §II we introduce the background of the connection set-up process and define our metrics to measure the time spent on connection set-up process. In §III we break down the connection set-up process into different sub-phases to find out the main contributors of the connection set-up time cost. In §IV we conduct the correlation analysis to see how certain factor affects the connection set-up process. §V gives case studies on mobile device and AP models. In §VI, we use the machine learning method to help mobile users select the APs. §VII discusses the related work and §VIII concludes the whole paper.

## II. CONNECTION TIME COST

In this section, we give the definition of the WiFi connection set-up time cost and sub-phases of the overall connection set-up process; then we present how we use the WiFi Manager App to record the time costs of the sub-phases.

The **WiFi Connection Set-up Time Cost** we use in this paper is defined as *the time span between the time a user clicks the SSID (service set identifier) name of the AP s/he wishes to connect and the time his/her device obtains the IP address*, which is shown in Figure 1. The WiFi connection set-up time cost is also abbreviated as **connection time cost** hereinafter. It is noteworthy that obtaining IP address does not guarantee the access to Internet. For instance, some public APs have portal webpages which require users to identify themselves. Such time cost is not studied in this paper. We present the sub-phases as below.

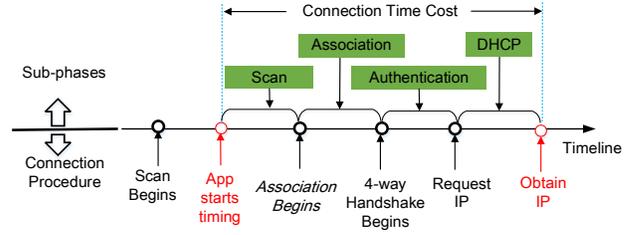

Fig. 1: Timeline of network activities when a mobile device connects to a WiFi AP.

*Scan*: The main purpose of scan is to update/confirm the available WiFi SSID list around the user's device. There are two types of scan: active scan and passive scan [3]. The active scan is triggered periodically: the mobile device first broadcasts the probe requests, and surrounding APs after hearing the probe requests will reply probe response packets, containing information such as the supporting physical rate. The mobile device will add the SSID of the AP into the candidate list, if it finds that the AP is compatible. In the passive scan, the list of available SSIDs can be updated by *beacon* packets broadcasted by APs periodically, *e.g.*, every 100ms [4].

*Association*: This is the must step for every connection set-up process, which has the following protocol: authentication request, authentication response, association request and association response. These 4 packets are sequentially sent and received. After the mobile device receives the association response, the WiFi MAC-layer connection is setup and user can send any MAC-Layer packets to the AP and vice versa. The authentication request/response is actually an empty authentication step, which is the legacy of previous WEP standard [5]. If the WiFi AP enables new encryption mechanism, *e.g.*, WPA2, the following real authentication phase is required.

*Authentication*: This is an extra phase for the APs, which is common in today's WiFi environment. The authentication consists of 4 MAC-Layer packets exchange. Note that a user may input a wrong password causing connection failure in this phase. We record such failures (as illustrated in Figure 2) for later study.

*DHCP*: In this phase, the mobile device will interact with the DHCP server, which can be either a standalone server or deployed on the AP. As soon as the device obtains the IP address, our App will mark this as the end of the connection attempt and calculate the *connection time cost*.

## III. BREAKDOWN OF WI-FI SET-UP PROCESS

In this part, we use our App to record the time cost of each sub-phase. Considering the extra computation and storage burden caused by timing each state transition, a version of our WiFi Manager App equipped with WiFi association breakdown is only deployed to $12,472$ carefully selected mobile devices, generating 706K connection attempts.

### A. WiFi Association: Success vs. Failure

A connection set-up process may end up with the following results, as illustrated in in Figure 2. The "Success" in Figure 2 represents those connection set-up processes which successfully obtain IP addresses within 30 seconds. "Timeout" represents those connection set-up processes which have not entered the *DHCP* phase within 30 seconds—the threshold that only $0.01\%$ connection set-up processes succeed eventually. "DHCP Failure" represents those connection set-up processes which entered *DHCP* phase but have not successfully obtained IP addresses. The rest results are "Wrong password" (*i.e.*, users typed the wrong password), "Switching to another WiFi" (*i.e.*, users decide to switch to another WiFi AP), clicking the "Forget WiFi" button, and "Switching off the WiFi interfaces". Also, less than $5\%$ of connection attempts are labeled as "Unknown". All these results indicate the termination of connection set-up processes.

In Figure 2, we observe that $45\%$ of connection set-up processes fail. The most common failure reasons are timeout and DHCP failure, which occupy about $24\%$ of overall connection attempts. The rest four types of failures in Figure 2 happen because users are not willing to connect to the current WiFi, *e.g.*, switching to another WiFi during connection or switching off the WiFi interface. In other words, we only study connections mobile users are willing to connect, including the following results: Success, Timeout and DHCP Failure.

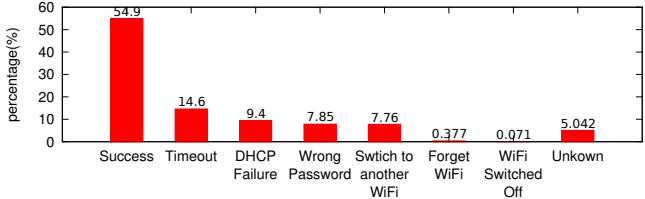

Fig. 2: Proportion of the different types of connection set-up processes.

### B. Distribution of Connection Time Costs

As explained in §II, a connection set-up process is successful if and only if the mobile device obtains the IP address within 30 seconds. We show the distribution of *connection time costs* of successful connection set-up processes in Figure 3. We observe that for about $80\%$ of those successful connection set-up processes, their *connection time costs* are smaller than 5 seconds. However, there is still a tail: about $3\%$ connection set-up processes have *connection time costs* larger than 15 seconds at last.

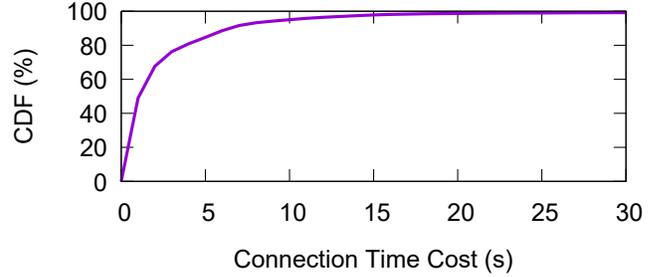

Fig. 3: The CDF of *connection time costs* for successful connection set-up processes.

### C. Breaking Down the Connection Set-up Process

The main advantage of breaking down is that it can easily find out what is the main contributor to the overall *connection time cost* using the time proportion of sub-phases. However, many state-of-the-art studies (*e.g.*, [2]) lack such detailed measurements of WiFi connections. In this part, we study the cumulative distribution function (CDF) of the absolute values of sub-phases in different Wi-Fi usage scenarios.

We divide the connection set-up processes into three categories according to their *connection time costs*: 0s∼7s, 7s∼15s and 15s∼30s. The CDF of each sub-phase's proportion is shown in Figure 4. The most obvious result is that the *association* and *authentication* sub-phases do not take too much time, compared to *scan* and *DHCP* sub-phases. It is reasonable because these two sub-phases consist of fixed number of WiFi packets exchange.

The second observation is that connection set-up processes with 0s∼7s and 7s∼15s *connection time costs* show similar distribution; while the connection set-up processes with 15s∼30s *connection time costs* show different patterns. For half of the connection set-up processes whose *connection time costs* are below 15 seconds, the *DHCP* phase occupies more than $80\%$, which is consistent with the conclusions in [2].

Surprisingly, for connection set-up processes whose *connection time costs* are larger than 15 seconds, the *scan* phase consumes more time than the *DHCP* phase, as illustrated in Figure 4a and Figure 4d. The time costs of *scan* phases for different connection set-up processes are shown in Figure 5. For connection set-up processes with *connection time costs* in 0s∼7s, $99\%$ of the scan time costs are smaller than 3.4s. For 7s∼15s class, $99\%$ of the scan time costs are smaller than 11.6s. In contrary, for the connection set-up processes of the 15s∼30s class, about $40\%$ of the scan time costs are larger than 11.6s.

To further explore why the time cost of the *scan* phase is large, we show the Android network state transition of connection set-up processes whose *connection time costs* are larger than 15s in Figure 6. The numbers beside the arrow show the total number of transitions happened in this dataset. Figure 6 shows that there are anomalous state transitions to *Disconnected* state. *Disconnected* state will trigger the re-connecting of the connection set-up process which begins with *Scanning* state. The scan time cost is large because many connection set-up processes re-enter the *Scanning* state multiple times.

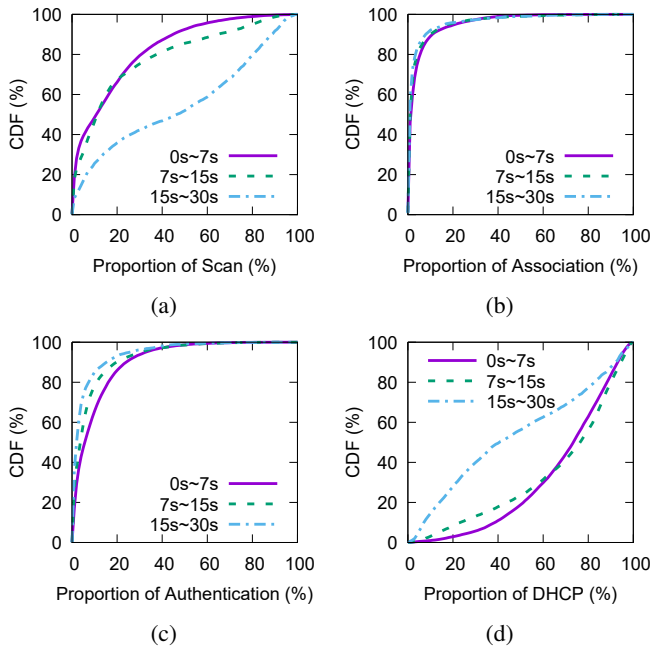

(a) (b) (c) (d)

Fig. 4: The distribution of the proportion for different sub-phases.

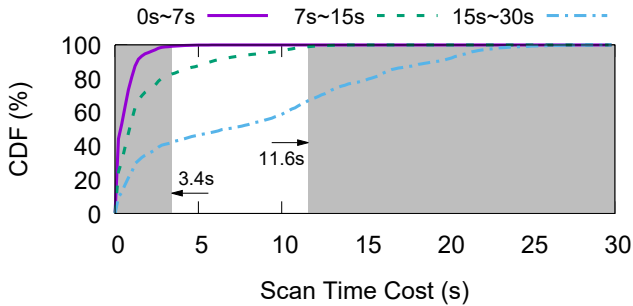

Fig. 5: The scan time costs of different kinds of connection set-up processes whose *connection time costs* are in 0s∼7s, 7s∼15s and 15s∼30s.

The state *Scanning* transits to *Disconnected* because there is no response for mobile device' probe response. The state *Associating* transits to *Disconnected* because the AP does not reply with the *Association Response* packets. There are two potential reasons: the request packets of mobile devices are not heard by the APs or the reply packets of APs are not heard by the mobile devices. However, all the connection set-up processes succeed eventually, which means that there are uncertainties for the receiving of WiFi packets, *i.e.*, WiFi packet loss. These kinds of packet loss are caused by the following reasons: variation of RSSI of APs measured at mobile devices, WiFi interference or system (APs or mobile devices) processing delay caused timeout. In the later part of the paper, we propose a solution to help the mobile devices choose better APs by considering all above factors together.

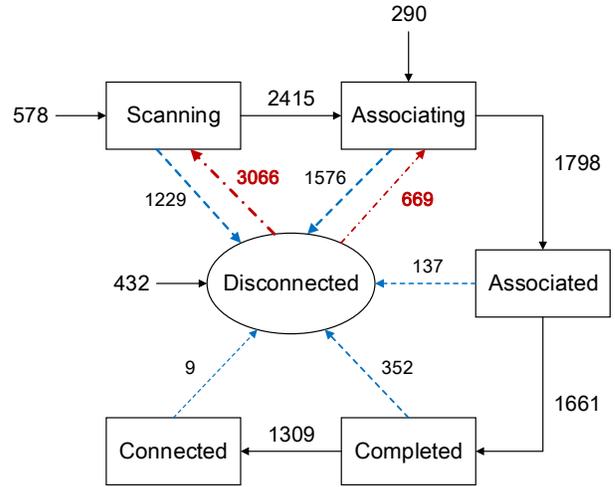

Fig. 6: The state transitions of connection set-up processes whose *connection time costs* are larger than 15 seconds.

### D. Summary

Here we highlight some conclusions from the above analysis.

First, only 54.9% connection set-up processes succeed. About 24% connection set-up processes fail to obtain the IP addresses eventually. For those successful connection set-up processes, 15% (5%) of them spend more than 5 seconds (10 seconds).

Second, for those 3% successful connection set-up processes whose *connection time costs* are larger than 15s, unexpected packet loss is the main culprit, which may be caused by variance of signal strength, WiFi interference and high system processing delay in APs or mobile devices. How to effectively take all these into consideration and help mobile devices to reduce the *connection time costs* are challenging.

## IV. CORRELATION ANALYSIS

In this section, we first give the brief introduction of the *connection log dataset* we collected, which is the foundation of our following analysis. Then we use both qualitative (**x-y** visualization) and quantitative (Relative Information Gain) methods to analyze how certain factor collected in the dataset affects the *connection time cost*.

### A. Introduction of Connection Log Dataset

We collect connection log at every mobile device equipped with our "WiFi Manager". We continuously collect one week data from May 3 to May 9 and log each connection set-up process to our central server. Thanks to the high popularity of our App, the one week dataset covers about 7 million unique APs and 5 million unique mobile devices in 4 different cities. The overall number of connection set-up processes reaches 0.4 billion. Each log in this dataset consists of two parts, which are listed in the top and bottom part of Table I. The top part of Table I is the environment related data and the bottom part of Table I is the performance related data. The environment related data is used to profile the context when each connection set-up process happens, which is helpful to find the root cause of the large connection time cost event. The performance data mainly contains the connection result and the *connection time cost* if the connection set-up process succeeds eventually. For the connection result in this *Connection Log Dataset*, we can only distinguish *success, wrong password caused failure* and *network caused failure*.

TABLE I: Fields of *connection log dataset* we used in this paper. The top part contains the features to profile the context when the connection set-up process happens. The bottom part indicates whether the connection set-up process succeeds and the corresponding *connection time cost* if succeeds.

| Abbreviation | Features | Description |
|---|---|---|
| *hour of day* | Hour of day. | Which hour the connection event happens in 24 hours. |
| *RSSI* | Received Signal Strength Indicator. | The signal strength of AP measured on the mobile device. |
| *number of devices* | Number of associated devices. | Number of devices currently associated on the AP. |
| *mobile device model* | Mobile device model. | The extracted information from the first eight characters of IMEI. |
| *AP model* | AP model. | The extracted information from the first eight characters of AP's BSSID. |
| *Encrypted* | Encryption type of the AP. | Whether the AP is encrypted using the password or not, *e.g.*, WPA2. |
| *IsPublic* | Is public AP? | The labeling result of an AP to decide whether the AP is public or not. |
| *result* | Connection result reported by the App. | Whether the App user successfully connects to the AP or not. |
| *connection time cost* | Connection time cost. | The time cost of the connection set-up process. |

Having this *Connection Log Dataset*, we first check some basic properties of this dataset in the dimension of time, location and users. We find that 80% of connection attempts are generated by 20% of users and happen in 20% of locations, which are consistent with Pareto Principle. The connection attempts are uniformly distributed in every day of one week and slightly vary in different hours of day, which reflect the schedule of people in one day.

We revisit the proportion of failure events (§III-A) and the distribution of *connection time cost* (§III-B) using *Connection Log Dataset* and get the similar results. Besides the overall distribution, we also care about how often a given user experiences the connection failure events. We calculate the proportion of failure events for one certain user. The results show that the proportion of failure events vary from 0 to 100% for different users. For less than 20% of users, the proportion of failure events stays under 0.8. We also check the fraction of users who have experienced connection failures across different time and locations. We find that this number changes smoothly with time and locations. In summary, the *connection failure events* are not caused by a small fraction of users in a few locations or a few days. Thus we can not reduce the number of connection failure events or long connection time cost events by simply using case by case analyses.

### B. Qualitative Analysis

To understand how each feature listed in Table I affects the *connection time cost*, we use the **x-y** visualization to show the variance of *connection time cost* with each feature. As there are thousands of different mobile device models and AP models, we omit the **x-y** visualization results for feature *mobile device model* and *AP model*. As an alternative, we will use qualitative analysis and detailed analysis in §V to study the impacts of *AP model* and *mobile device model* to the *connection time cost*. We bin the non-category features using different bin-widths: 5dBm for *RSSI*, 10 for *number of devices*. Then we calculate the mean *connection time cost* for each bin, which is shown in Figure 7.

**Association timing affects the *connection time cost*.** From Figure 7a we can see that the *hour of day* affects the *connection time cost*. Figure 7a shows that the *connection time costs* in the daytime (7am to 7pm) are larger than the *connection time costs* at night. This is mainly because that peoples activities are different at different hours, *e.g.*, the APs in a restaurant will have peak number of associating attempts at the lunch or dinner hour.

**Connections with higher RSSI tend to have smaller average *connection time costs*.** There is clear monotonous relationship between the *RSSI* and *connection time cost*. The *connection time cost* varies from 5 seconds to 2.5 seconds when *RSSI* increase from -90dBm to -60dBm. It is noteworthy that in the latest Android devices, there is a maximum *RSSI* value, *i.e.*, *RSSI* ≥-55dBm will be recorded as -55dBm.

### C. Quantitative Analysis

In this section, we mainly use Relative Information Gain [6] and Kendall coefficient [7] to quantitatively evaluate the relationship between the features and *connection time cost*, which is inspired by [8], [9]. The values are shown in Table II. To help explain the results, we give a brief introduction of these two metrics.

*Relative Information Gain* (RIG hereinafter): For the raw $(\mathbf{x}, \mathbf{y})$ input, RIG is used to show the help of knowing $\mathbf{x}$ to predict $\mathbf{y}$. In our calculation, we bin $\mathbf{x}$ and $\mathbf{y}$ vectors according to their characteristics. The bin-width we use for *hour of day*, *number of devices* and *RSSI* ($\mathbf{x}$) is 1, 10 and 5dBm, respectively. For *connection time cost* ($\mathbf{y}$), we use 100ms as bin-width.

*Kendall coefficient* (Kendall hereinafter): Kendall is a rank correlation calculation. Intuitively, the Kendall coefficient between two variables $(\mathbf{x}, \mathbf{y})$ is high when they have similar rank. We first bin the vector $\mathbf{x}$ using the same bin-width as information gain and calculate the mean value of $\mathbf{y}$ which falls into each bin. Then we calculate the Kendall coefficient between each **x** and **y**. We do not calculate Kendall for categorical features because Kendall is mainly used to show the monotonicity between certain feature and *connection time cost*, which is not suitable for categorical features.

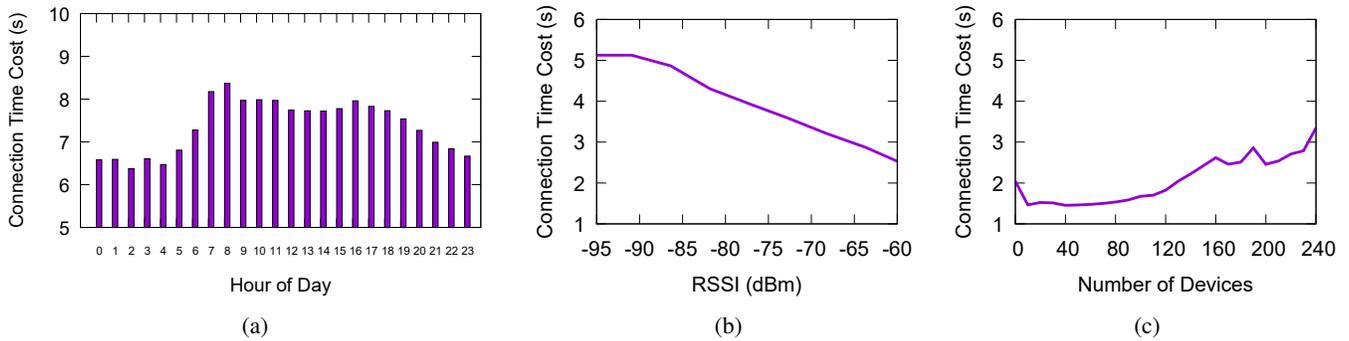

Fig. 7: Correlation between different features and *connection time cost*. -60dBm in *RSSI* represents the [-60dBm, -55dBm] bin. 10 in *number of devices* represents the [10,20] bin.

TABLE II: RIG and Kendall between different features and *connection time cost*.

| Features | RIG | Kendall |
|---|---|---|
| mobile device model | 0.156 | / |
| AP model | 0.078 | / |
| RSSI | 0.020 | -0.395 |
| number of devices | 0.006 | 0.208 |
| hour of day | 0.005 | / |

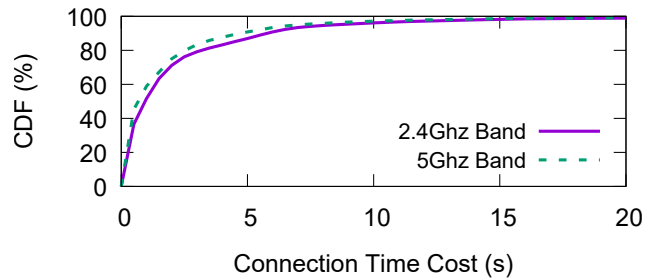

Fig. 8: The CDF of *connection time cost* under different WiFi bands.

**Mobile device model and AP model have highest RIG**. High RIG indicates that knowing the model of device/AP has great help to predict the *connection time cost*. For instance, our measurement shows that the average *connection time costs* of all HTC phones is 1.3× larger than that of Samsung, the reasons might be that some of the device models of Samsung have equipped with top-level hardware (*e.g.*, chipset) which always means few design bugs and faster response time.

**RSSI has large RIG and the highest Kendall**. It is natural that *RSSI* has highest Kendall because the *connection time cost* has clear monotone decreasing relationship with *RSSI*, which is shown in Figure 7b. *RSSI* also has acceptable RIG to *connection time cost* and that explains why many existing methods choose APs via signal strength [10].

**Number of devices helps little**. Table II shows that *number of devices* has low RIG and Kendall coefficient. It is confirmed by Figure 7c because there is no clear correlation between the *connection time costs* and the *number of devices*. The main reason is that each AP has its capacity to hold certain number of devices and the performance begins to degrade when the number of devices exceeds this threshold. If the number of devices stays under that threshold, its impact on *connection time cost* is quite small. In Figure 7c, this threshold stays around 100.

It is noteworthy that our feature *number of devices* represents the number of devices currently associated with my AP, which is different from the *number of surrounding devices* in [11]. *Number of (associated) devices* affects the performance in terms of local contention. The larger the number of associated devices is, the higher possibilities of local contention are, which further leads to longer local queue. On the other side, [11] uses the overheard wireless packets to infer the *number of surrounding devices*. *Number of surrounding devices* reflects the device density of the surrounding environment which affects the contention on the same channel.

### D. Other Features

**2.4GHz Band vs. 5GHz Band:** Besides the selected features listed in Table I, we study the differences of *connection time cost* under different WiFi bands in Figure 8, *i.e.*, 2.4GHz band and 5GHz band. The information of WiFi band is inferred from the SSID of certain connection attempt. For example, if the SSID is "Tsinghua-5G", we can infer all the connection attempts to this SSID are under 5GHz band. Note that only a small fraction of SSIDs are named in this pattern, but thanks to the large dataset, we have enough data to profile the connection attempts under different WiFi bands in Figure 8. The data summary of Figure 8 is shown in Table III.

**Roaming:** Roaming is an important feature under the enterprise networks. Different customized enterprise networks may have different roaming configurations. However, roaming rarely happens under home networks. To what extent the roaming under enterprise network affects the *connection time*

TABLE III: The summary of the data which can distinguish which band the connection attempt happens.

|  | #SSIDs | #connections | #Users | #Failure Events |
|---|---|---|---|---|
| 2.4GHz | 451 | 80874 | 7921 | 22141 |
| 5GHz | 451 | 21979 | 1246 | 5370 |

*cost* is an interesting topic. In enterprise networks, there are two kinds of connection set-up processes. For those wireless stations which first come into the network, they perform the initial connection set-up processes. The other kind is roaming, which happens when a station is already associated with one AP of the enterprise network and wants to connect to a new AP with the same SSID name due to the mobility of the user.

The initial connection set-up process of enterprise network is a little different with the home network because of the existence of IEEE 802.1X EAP (Extensible Authentication Protocol) process in enterprise network, which will cost extra $2 * N_e(T_w + T_a) + T_a$ time. $N_e$ is the number of round trips needed by EAP, which equals 4 in common cases. $T_w$ is the one-way WiFi hop latency. $T_a$ is the wired latency between the AP and AAA (Authentication, Authorization, and Accounting) server [12]. However in home network, the pre-shared-key is used, *i.e.*, no EAP process [13]. The measurement results of these differences are discussed in §V-B.

There are two common fast-handover (roaming) methods in enterprise networks: CAPWAP and HOKEY [12]. Their main goals are reducing the extra time costs imported by EAP processes, which are not included in home networks. Furthermore, CAPAP and HOKEY still need the discovery phase, same with the processes in home networks. This is confirmed by our real world measurements. [12] concludes that the maximum extra time of roaming in enterprise network is a wireless round-trip time compared with the connection attempt in home network. The CDF results show that the *connection time cost* of roaming in enterprise networks and *connection time cost* in home network are almost the same.

## V. CHARACTERIZING THE IMPACTS OF DIFFERENT MOBILE DEVICE AND AP MODELS

In this section, we will go one step further to discuss how the AP model and mobile device model affect the *connection time cost*. Using these detailed analyses, we want to show that there is extra information in these two features, *AP model* and *mobile device model*. This is confirmed in the analyses of [14] and [15]. If we can learn these information in advance from collected dataset, there will be great help to predict the *connection time cost* for mobile devices.

### A. How the mobile device model affects the connection time cost

There are evidences (Table II) that *mobile device model* has high RIG on the *connection time cost*. We try to dig out the hidden information provided by the *mobile device model*, including the mobile operating system, chipset, CPU frequency, RAM size, etc. We sort the mobile device models according to their average *connection time costs* and select some popular models shown in Table IV. Each mobile device model contains no less than 0.5 million pieces of data. The average *connection time costs* are calculated by using the connection set-up processes whose RSSI is higher than -60dBm to exclude the impact of low signal strength. Under each class of device models, there are thousands of different users connecting to thousands of unique APs. The connection attempts happen across all days in one week and thousands of different places. By averaging the connection time cost in various environments, the temporal and spatial influences are alleviated so that we can focus on the effects of only the device models on *connection time cost*. From Table IV we can draw the conclusions as follows.

- The chipset can affect the *connection time cost*. Here we take the MEIZU M1 Note and MEIZU M2 Note as an example. From Table IV we can see that these two mobile devices are from the same company and have the same operating system and RAM size. The chipsets for MEIZU M1 Note and MEIZU M2 Note are MediaTek 6752 and MediaTek 6753 respectively. MediaTek 6753 and 6752 use the same architecture: ARM Cortex-A53. The WiFi module on both chipsets is also the same: MT6625L, which means that the detailed specifications of wireless interfaces are the same, *e.g.*, antennas, filters. The big difference is the CPU frequency. The CPU frequency of MediaTek 6753 is 1.3GHz and the CPU frequency of MediaTek 6752 is 1.7GHz. We draw the CDF of *connection time cost* in Figure 9a, which shows great differences in *connection time costs* between these two phones.

- The operating system matters under the same hardware equipment. We take SAMSUNG G9280 and MEIZU PRO 5 as an example. We can see in Table IV that the hardware configurations are almost the same. They all use the same chipset (Exynos 7420) together with the same CPU and RAM. The only difference is the operating system. The CDF of *connection time costs* for two models are shown in Figure 9b. There are many MEIZU PRO 5 users complaining about the performance of MEIZU PRO 5 in 5GHz band. To further figure out the problem, people install "WiFi Analyzer" on two phones to scan the surrounding 5GHz APs. The MEIZU PRO 5 can only report one 5GHz AP whose received signal strength is -80dBm. However, the signal strength of this AP reported by SAMSUNG G9280 is -60dBm. Moreover, the SAMSUNG G9280 can discover more surrounding APs in 5GHz band. This may be caused by the wrong calibration of the signal strength for 5GHz packets on MEIZU PRO 5 which leads the system to drop most of the 5GHz packets because of the low signal strength problem.

### B. How the AP model affects the connection time cost

*1) Public AP vs. Private AP:* During our investigation, we find that there is a feature which can effectively distinguish

TABLE IV: The software and hardware parameters for different kinds of mobile device models.

| Average connection time cost | Device model | Operating System | Chipset | CPU Frequency | RAM Size | Wireless Interface |
|---|---|---|---|---|---|---|
| 475ms | **MEIZU M1 Note** | Flyme | MediaTek 6752 | 1.7GHz | 2GB | IEEE a/b/g/n |
| 754ms | **SAMSUNG G9280** | Android OS | Exynos 7420 | 2.1GHz | 4GB | IEEE a/b/g/n/ac |
| ... | ... | ... | ... | ... | ... | ... |
| 2463ms | **MEIZU M2 Note** | Flyme | MetiaTek 6753 | 1.3GHz | 2GB | IEEE a/b/g/n |
| 3534ms | **MEIZU PRO 5** | Flyme | Exynos 7420 | 2.1GHz | 4GB | IEEE a/b/g/n/ac |

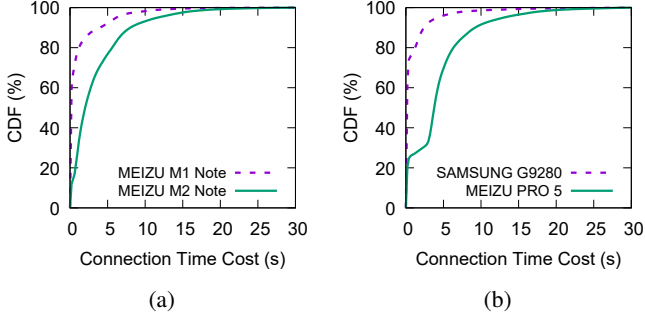

Fig. 9: (a) shows the frequency of CPU can affect the *connection time cost*. (b) shows the operating system can affect the *connection time cost*.

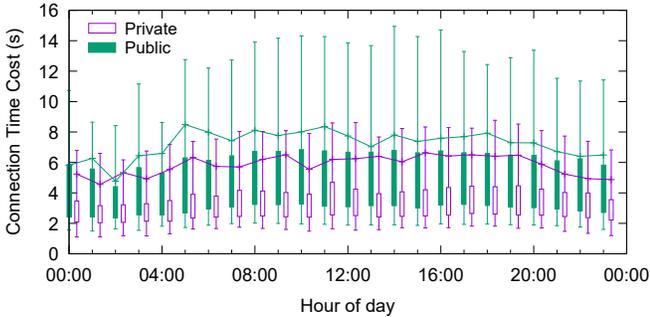

Fig. 10: The *connection time cost* for public and private APs in one day (May 3). We calculate the minimum, 25, 75 and 90 percentile of the *connection time costs* for public APs and private APs in each hour. The dotted lines show the average *connection time cost* respectively.

APs in terms of *connection time cost*. This feature is *IsPublic* in Table I. If *IsPublic* is TRUE, it means the AP is a public AP, otherwise, this AP is a private AP. Here we give the definition of the *private* and *public*.

- Private APs: APs which provide private WiFi services for a relatively small number of users.
- Public APs: APs which provide public/open WiFi services. These APs often have a larger number of users associated with them and there are large variances on the devices associated with them.

To see the differences of the *connection time costs* between public and private APs, we manually label 200K APs as public and private APs and draw the *connection time costs* distribution of these two kinds of APs in Figure 10. From Figure 10, we can conclude that the *connection time costs* of public APs in one day are consistently larger than private APs. Figure 10 also shows that the 90% *connection time costs* of private APs are consistently smaller than 10 seconds. This observation tells us that whether the AP is public AP or not is a good feature to predict the *connection time cost* but how to accurately label the APs into public and private is not easy. In the following we will address this problem.

*2) How can public or private characteristic help connection time cost prediction:* Intuitively, we find that some AP models tend to be deployed to provide only one kind of WiFi service (private or public). For instance, many enterprises use the Cisco wireless solution which contains limited kinds of AP models. To confirm this, we group our manually labeled 200K APs using their AP models. Totally we get 2802 distinct AP models. Our statistic results show that 27% of AP models are only used as public APs. 32% of AP models are only used as private APs. This means that if we learn this information of AP models in advance, we can effectively classify 59% APs into public and private using their models. This finding gives us the motivation to add the *AP model* as a feature into the machine learning model to classify the *connection time cost* in §VI.

## VI. MODELING

In the previous sections, we studied the importance of different features and know some reasons to answer the question like "why my connection time is so long?". In the following parts, we will focus on "what can I do to reduce the *connection time cost*?". First we will train a machine learning model to predict the *connection time cost* using the features listed in Table I. Then we propose an AP selection algorithm which uses the machine learning model. At last, we evaluate the effectiveness of our algorithm in our collected dataset.

### A. Machine Learning Model

**Model Selection**: Though the decision tree model has interpretability and intuition, the accuracy of single tree is not satisfactory. The random forest works by constructing multiple decision trees using random selected features and the output class for the classification is the mode of output classes of all the individual trees. It is an ensemble of decision tree, thus it performs better than single decision tree in most of the cases. So we choose random forest model to predict the *connection time cost* for each connection attempt. The threshold we use to label the connection attempt into FAST or SLOW class is 15 seconds. If the *connection time cost* of a connection attempt is larger than 15 seconds, the connection attempt will be labeled as SLOW and vice versa. The threshold is decided according to the breaking down result (see Figure 4). The connection set-up processes whose *connection time costs* are larger than 15 seconds or smaller than 15 seconds show great differences.

TABLE V: Accuracy of random forest model. The parameters we use for this model are: Tree depth=90, #Tree=100, weight=0.3.

| Label. | Precision | Recall | Features Used |
|---|---|---|---|
| FAST | **0.91** | 0.49 | *hour of day, RSSI, AP model,* |
| SLOW | 0.48 | **0.90** | *mobile device model, Encrypted* |

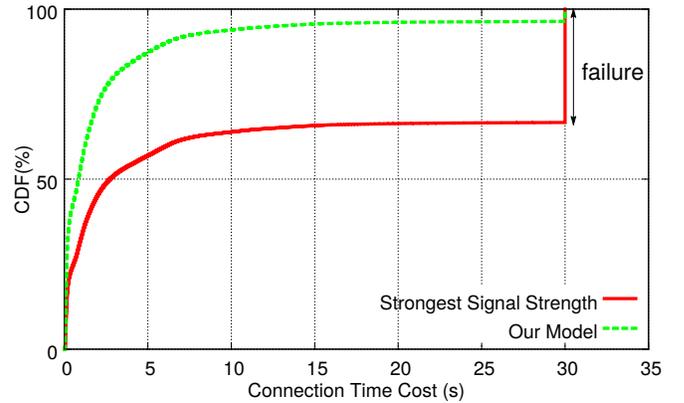

Fig. 11: Comparison of AP selection algorithms.

**Feature Selection**: There are two criterions for feature selection: (1) All the features we use should be easily measured by mobile devices. (2) Use as few features as possible under acceptable accuracy. The final features we choose to train the *connection time cost* includes *hour of day, RSSI, mobile device model, AP model, Encrypted* in Table I. According to the analysis in §V-B2, we use *AP model* to replace *IsPublic* because the feature *IsPublic* can be inferred from the *AP model* which can be easily obtained by the mobile devices. To further make the model more friendly to use, we exclude *number of devices* from the feature set for two reasons: (1) *number of devices* is the AP-side information and is hard to obtain on the mobile devices. (2) After removing the *number of devices* from the feature set, our random forest model still has acceptable accuracy. The accuracy of the random forest model we finally choose for the next section is shown in Table V.

### B. Our Machine Learning Based AP Selection Algorithm

In this part, we use our machine learning model trained before to help mobile users choose APs in order to reduce the *connection time cost*. Our algorithm can be summarized as the following two steps.

Firstly, for each AP in the mobile device's candidate list, we collect the features listed in Table V and input these features to the machine learning model we trained before. If the classification result is SLOW, we will add this AP into SLOW set and vice versa. At the end of this step, we will divide the candidate APs into SLOW and FAST sets. Then we choose the AP with the strongest signal strength from the FAST set for the mobile device to connect to and our algorithm ends.

### C. Evaluation

In this section, we use our collected data to evaluate the performance of our algorithm. The baseline algorithm we used to compare with is called "Strongest Signal Strength". This algorithm help mobile users choose APs which have strongest signal strength in each mobile device's candidate list. This method is quite simple but has great popularity which is used by many state-of-the-art works.

To evaluate, we first divide our *connection log dataset* into two parts, each subset contains 50% of the overall data. For these two parts, one is used for tuning the model (including training and validation), the other is used for evaluating in this section. Note that the evaluation dataset is never seen by our machine learning model before. This fresh dataset ensures that we can accurately evaluate the performance of our algorithm if we deploy our algorithm in the wild, where many of the APs will not be seen by the mobile devices before. Then we run the baseline algorithm and our algorithm on the evaluation dataset to help mobile devices in this dataset select APs. Each algorithm will output an AP list contain the APs it chooses for each mobile device in the dataset. As we already know the real *connection time costs* when connecting to these APs, we can evaluate our algorithm by comparing these two output files.

The distribution of the *connection time costs* after using these two algorithms are shown in Figure 11. We can see that if all the mobile devices in our dataset equipped with the "Strongest Signal Strength" algorithm, there are about 33% connection attempts fail. We record the *connection time cost* of connection failure attempts as 30 seconds in Figure 11. As a comparison, if all the mobile devices in our dataset are equipped with our algorithm, the machine learning based AP selection algorithm which is explained before, there are less than 3.6% connection attempts fail. The 80% time costs is only 3 seconds, compared to more than 30 seconds using the baseline algorithm, which is 10× reduction on the 80% *connection time cost*. The reason why our algorithm perform better is that the baseline algorithm only chooses APs by the signal strength. However, there are large possibilities of connection failure events even the measured signal strength is highest at the mobile devices. Our model can predict these connection failure events in advance with a high accuracy and avoid the mobile devices from connecting to the SLOW APs.

However, our algorithm has limitation. We will reduce the number of available APs (*i.e.*, APs in the FAST set) for mobile users to connect to. This is because some of the APs whose *connection time costs* are smaller than 15 seconds can be classified into the SLOW set and there will be fewer available APs in the FAST set for mobile users to associate to. Thus there are trade-offs between the performance gain and number of available APs to associate. Here we define a metric called "Proportion of available APs" (PoA) as $\frac{|FAST\ set|}{|FAST\ set|+|SLOW\ set|}$. The PoA indicates the fraction of FAST APs after classified by our model. We list different models and their corresponding PoA metrics in Table VI. We can see that the PoA is $0.33$ when the random forest's Recall(SLOW) is $0.9$. This PoA value means our model can guarantee the mobile users have at least one AP to associate when there are more than 3 candidate APs around the mobile users. It is acceptable because recent surveys [16], [17] show that there are usually tens of APs one can associate in representative urban area.

TABLE VI: The corresponding proportion of available APs (PoA) which mobile users can associate with after using our model under different Recall(SLOW).

| Precision(SLOW) | Recall(SLOW) | PoA |
|---|---|---|
| 0.40 | 0.98 | 0.15 |
| 0.43 | 0.96 | 0.21 |
| 0.49 | **0.90** | **0.33** |
| 0.54 | 0.84 | 0.40 |

Our model is trained using all the log traces collected in one month and used for the next month to predict the *connection time cost*. We only need to train the model once at the beginning of the month. We assume that our dataset is large enough which can cover almost all the possible parameter combinations. How to incrementally add the newly collected traces in last month into the existing model with a proper weight is an interesting topic and will be left as future work. As our model is trained offline, the only burden imported by our algorithm is the exchanging time of several packets (no more than 10 in most cases), *i.e.*, to send the candidate list collected from the mobile devices to the remote server and receive the filtered list back from the server. As these packets are pipelined, one round-trip time plus the packets transmission time are enough. The mobile devices use the alternative methods, *e.g.*, cellular network, to communicate with the remote server before the successful WiFi association, the average *round-trip time* is millisecond level. In this way, the extra time imported by our algorithm is negligible compared with the overall *connection time cost*.

## VII. RELATED WORK

There are plenty of works which focus on the WiFi performance measurement [18]–[21]. [18] aims to estimate the available throughput for certain AP-Client link. [21] explores the latency at the AP-side. However, based on our previous measurement, there is urgent need to pay attention to the connection set-up time cost metric because high connection failure rate has already affected the user experiences.

There are few works focusing on the connection set-up process. Most of the current works which involve the connection set-up process are about WiFi handover mechanism: [22]–[29] aim to reduce the handoff latency. The connection set-up process studied in this paper is different from the handover process because the handover process contains discovering the next AP while associating on the prior AP.

[2] is the first work focusing on the WiFi connection set-up process. Suranga Seneviratne et.al collect data from 13 voluntary mobile devices and draw some observations. However, the distribution of connection set-up time cost in the wild still remains unknown. Besides, some of the observations may be different when we use large scale measurement. For instance, we find that when the connection set-up time cost is larger than 15 seconds, *scan* phase is the main contributor, instead of *DHCP* phase mentioned by [2]. [25] studies the different phases of connection set-up process but is under vehicular scenario.

## VIII. CONCLUSION

Though most of the state-of-the-art works focus on WiFi latency and throughput, we want to draw people's attention to the WiFi connection set-up time cost using our measurement and analysis. Using a mobile WiFi Manager App which tops the Android/iOS App market, we continuously collected $0.4$ billion WiFi connection attempts. The results show that $45\%$ of the WiFi connection attempts fail and about $5\%$ of attempts consume more than $10$ seconds.

To further understand the WiFi connection set-up process, we break down the overall process into four sub-phases and find the *scan* phase is the main culprit for the large connection set-up time cost event. The correlation analysis finds that though the signal strength is important, knowing the AP model and mobile device model has great help to predict the connection set-up time cost. To the best of our knowledge, we are the first to add AP model and mobile device model as features which greatly increases the accuracy to predict the connection set-up time cost.

Based on the comprehensive measurement and detailed analysis, we propose a machine learning based AP selection algorithm. This algorithm classifies the candidate APs into the SLOW or FAST set by taking the features of APs as inputs of the machine learning model. Based on the classification results, our algorithm avoids the mobile device from connecting to those problematic APs which are classified into the SLOW set. The evaluation result shows that compared with the baseline algorithm which selects APs purely using the signal strength, we can reduce the connection failure from $33\%$ to $3.6\%$. The $80\%$ connection set-up time costs can be reduced by $10\times$.

## ACKNOWLEDGMENTS

We thank the anonymous reviewers for their valuable feedbacks. We are grateful to the members in Tencent WiFi Manager team for deploying the customized App. We thank Miao Tian and Juexing Liao for their helpful suggestions and proofreading. This work has been supported by National Natural Science Foundations of China (NSFC) under Grant 61472210 & 61472214 & 61402247, the Sate Key Program of National Science of China under Grant No.61233007, the National Key Basic Research Program of China (973 program) under Grant No.2013CB329105, the National High Technology Development program of China (863 program) under Grant No.2013AA013302, Tsinghua-Tencent Joint Laboratory for Internet Innovation Technology.